\documentclass{aa}
\usepackage{graphicx}
\usepackage{txfonts}
\begin{document}
   \title{Time dependent mixing in He-burning cores: the case of
          NGC1866}

   \author{P. Ventura\inst{1}
          \and
          M. Castellani\inst{1}
          }

   \offprints{P. Ventura}

   \institute{Osservatorio Astronomico di Roma
              Via Frascati 33 00040 MontePorzio Catone - Italy\\
              \email{ventura@mporzio.astro.it, m.castellani@mporzio.astro.it}
             }

   \date{Received ... ; accepted ...}

   \abstract{
   We examine the convective core helium burning phase of 
   intermediate 
   mass stars, and investigate the role of coupling of nuclear burning 
   and mixing on the extension and duration of the blue loop phase. 
   We compare the theoretical scenario with the distribution of stars
   in the colour-magnitude (CM) diagram of the Large Magellanic 
   Cloud (LMC) cluster NGC1866, whose largely populated clump of He-burning
   stars is equally populated in the blue and red side.
   We compare the distributions expected by adopting either 
   a diffusive scheme within the instability
   regions, in which nuclear burning and mixing are self-consistently coupled, or the traditional
   instantaneous mixing approximation.
   We analyze with particular care the sensitivity of the results to:
   a) the e-folding distance with which
   the velocity of convective eddies decays outside the formal border of 
   the convective regions fixed by the Schwarzschild criterion; 
   b) the convective model adopted to evaluate the temperature 
   gradient; c) the rate of reaction $^{12}C+\alpha \rightarrow ^{16}O$.  
   Models not including convective overshoot are also commented. 

   \keywords{Stars: evolution --
                Stars: interiors --
                Stars: Hertzprung Russell and C-M diagrams
               }
   }

   \maketitle
%

\section{Introduction}
The young stellar cluster NGC1866 in the Large Magellanic 
Cloud has been extensively studied in the last decades because 
it provides a unique opportunity to study the physics 
of the interiors of intermediate mass stars, particularly 
for what concerns the extension of the convective cores 
during the phases of both hydrogen and helium burning. 
It is young enough (Age $\leq \sim 200$Myr) to have 
turn-off masses of approximately $M\sim 4.5M_{\odot}$,
which are well known to develop extensive convective cores 
during their main sequence phase. At the same time 
the data available show an extraordinarily well populated clump
region (Testa et al. 1999; Brocato et al. 2003), 
with $\sim 100$ He-burning stars, 
and allow to compare the relative duration of 
the permanence of He-burning
stars in the red and blue parts of the CM diagram.

The observational data have been used up to now primarily to find out 
informations regarding the convective central regions during the 
phases of central burning (Becker \& Mathews 1983;
Brocato \& Castellani 1988;  Chiosi et al. 1989;  Brocato et al. 1994; 
Barmina et al. 2002; Brocato et al. 2003), and
particularly on the extension of the so-called overshooting region, 
i.e. the zone behind the formal border fixed by the Schwarzschild 
criterion in which the convective eddies are
allowed to travel due to inertia despite the opposition of the 
buoyancy forces (Cloutman \& Whitaker 1980; Xiong 1980; Canuto 
et al. 1996; Bressan et al. 1981).
 
The computations of detailed luminosity functions have provided a 
powerful tool to infer the relative duration of the phases of helium 
and hydrogen burning, which is directly correlated to the extension 
of the convective core during the MS phase, and the amount of
extra-mixing which must be taken into account. Unfortunately, at the 
moment no common agreement has yet been reached, and the photometric 
study of NGC1866 has been presented by different groups
either as a clue of a considerable amount of extra-mixing beyond 
the border of the instability regions (Barmina et al. 2002), or as 
a clear evidence that such a region is so narrow that overshooting 
can be completely neglected in the evolutionary calculations (Testa 
et al. 1999; Brocato et al. 2003). 
To understand the level of uncertainty, suffice it to say that
the differences in the ages of the cluster quoted in the literature 
amount to $\sim 50\%$, ranging from $\sim 100$Myr up to $\sim 150$Myr.

The idea behind this paper is not to provide another estimate of the 
age of the cluster through a comparison either between the 
theoretical isochrones and the observed CM diagram or
between the luminosity functions; rather, we focus our attention 
on the observed distribution of He burning stars in the CM diagram, 
in an attempt to reproduce the observed ratio between
blue and red clump stars ($B/R \sim 1$). 
We also discuss and compare the results obtained by varying the micro and 
macro-physics inputs adopted to calculate the evolutionary sequences.

Our main goal is to check the sensitivity of the global duration 
of the helium burning phase and of the time spent in the blue region 
of the CM diagram on the way nuclear burning and mixing of chemicals 
are coupled in the central convective regions during the two major phases
of nuclear burning. We compare the results obtained by assuming 
a diffusive approach, in which the two processes are self-consistently 
coupled, with the instantaneous mixing approximation, still largely 
adopted in most of evolutionary codes.

We use the observed distribution of clump stars in NGC1866 
as a test of the diffusive scheme for the masses involved;
we postpone to a forthcoming paper a more general discussion 
concerning a more extended mass range.

We complete our study by testing the sensitivity of the results also 
on the convective model adopted, our code allowing to calculate the 
temperature gradient within convectively unstable regions through 
the FST or the MLT models (see Sect.2.1), the rate of 
the reaction $^{12}C+\alpha \rightarrow ^{16}O$ and the e-folding 
distance of the convective velocity from the formal border
of the convective core, fixed by the Schwarzschild criterion.
 
A general description of the code ATON used to compute the 
evolutionary sequences is given in Sect.2. In Sect.3 we describe 
the main evolutionary phases of the typical intermediate mass
star which is likely to populate the clump of NGC1866.
A comparison between the results obtained by varying the scheme 
for mixing is presented in Sect.4. Sects.5 to 7 are focused 
on the effects of changing, respectively, the e-folding
distance of convective velocities decay within radiatively
stable regions, the convective model, and the cross-section
of the reaction $^{12}C+\alpha \rightarrow ^{16}O$.


\section{The evolutionary models}

The stellar models calculated in this paper have been obtained
by using the ATON2.0 code described by Ventura et al. (1998).
The interested reader may find on the 
afore mentioned paper a detailed description of the numerical 
structure of the code, and of the macro and micro-physics which is used to
simulate the stellar evolutions. Here we briefly recall the most 
important inputs adopted.

\subsection{The treatment of convection}
The calculation of the temperature gradient in stellar regions unstable 
to convection has been traditionally performed by adopting the Mixing 
Length Theory (MLT) (Vitense 1953; B\"ohm-Vitense 1958), which is
based on the simplified assumption that the convective flux is 
carried by eddies which first form and travel for an assumed 
distance (the mixing length), then lose their identity, transferring 
to the surrounding layers their energy content. 
Both the dimension of the afore mentioned eddies and the mixing length are 
assumed to be directly proportional
to the distance of exponential decay of pressure ($H_p$), with a constant 
of proportionality $\alpha_{MLT}$ which is commonly calculated in 
order to reproduce the evolution of the Sun, i.e. to fit the solar 
radius and luminosity at the present age of the Sun:  
we find  $\alpha_{MLT}=1.7$.

Alternatively, still in the framework of local convective models, 
in the last decade a new prescription was presented: the 
Full Spectrum of Turbulence (FST) model 
(Canuto \& Mazzitelli 1991; Canuto et al. 1996). 
The main differences with respect to MLT are that
all the spectrum of eddies' dimensions is taken into account, 
and the mixing length is assumed to be simply the distance to 
the closest convective border ($l=z$). This latter model provides 
convective fluxes which are approximately $\sim 10$ times larger 
than the MLT ones in regions of high convective efficiency, and 
sensibly lower in stellar zones where the efficiency of
convection is very poor, i.e. in proximity of the formal border 
or in the sub atmospheric layers. 

The code ATON2.0 can be easily managed in order to use either 
of the convective models.

\subsection{Nuclear network}
With respect to Ventura et al. (1998) the nuclear network has been 
extended and now includes 30 chemicals:$H$,$D$,$^3He$,$^4He$,$^7Be$,
$^7Li$,$^{12}C$,$^{13}C$,$^{14}N$,$^{15}N$,$^{16}O$,$^{17}O$,
$^{18}O$,$^{18}F$,$^{19}F$,$^{20}Ne$,$^{21}Ne$,
$^{22}Ne$,$^{22}Na$,$^{23}Na$,$^{24}Mg$,$^{25}Mg$,$^{26}Mg$,
$^{26}Al$,$^{27}Al$,$^{28}Si$,$^{29}Si$,$^{30}Si$,$^{31}P$,$n$. 

All the main reactions of the proton-proton chain, CNO, Ne-Na
and Mg-Al cycles, and of He burning have been considered, for a total
of 64 reactions. The relevant cross-sections are taken either 
by Caughlan \& Fowler (1988) or by Angulo et al (1999).

\subsection{Convective mixing and nuclear burning}
The extension and the borders of all the convective regions
are established by the Schwartzschild criterion.

The treatment of nuclear burning within convective regions is based 
traditionally on the instantaneous mixing approximation: it is assumed 
that convection is so fast in mixing the chemical species within the 
whole instability region that this latter can be assumed to be always 
homogenized. From a numerical point of view this latter assumption provides
a great simplification, because it is possible to calculate average 
chemical abundances and nuclear cross-sections, which are used for
all the mesh points included in the convective region.

Alternatively, it is necessary to solve for each element the diffusion 
equation (Cloutman \& Eoll 1976):

\begin{equation}
$$
  \left( {dX_i\over dt} \right)=\left( {\partial X_i\over \partial t}\right)_{nucl}+
  {\partial \over \partial m_r}\left[ (4\pi r^2 \rho)^2 D {\partial X_i
  \over \partial m_r}\right]  \label{diffeq}
$$
\end{equation}

\noindent
stating mass conservation of element $i$. The diffusion coefficient $D$ is

\begin{equation}
$$
   D=16\pi^2r^4\rho^2\tau^{-1}  \label{Ddef}
$$
\end{equation}

\noindent
The turbulent diffusion time scale, $\tau$, is related to the correlation 
between density and velocity according to:

\begin{equation}
$$
 <\rho_i'u_i'>=-\tau{\partial \rho_i\over \partial r} \label{taudef}
$$
\end{equation}

The knowledge of the second order momentum in eq.(\ref{taudef}) 
requires the solution of the Navier-Stokes equations, which is not 
available. Thus, it is customary to rely on a local approximation 
for D, that is

\begin{equation}
$$
D={1\over 3}ul
$$
\end{equation}
where $u$ is the convective velocity and $l$ is the convective scale 
length.

Careful details of the way of solving  eq. (\ref{diffeq}) can be found
in Ventura et al. (1998) (see, in particular, the Appendix and Sect. 2.3).

A diffusive approach was used by Deng et al.(1996a;b) and Salasnich 
et al.(1999) to study the evolution of massive stars of solar and LMC 
metallicity during their phases of hydrogen and helium burning.
 
The two schemes to treat nuclear burning within convective regions 
need also two different approaches to deal with overshooting. 
In the instantaneous mixing approximation it is simply assumed, 
for the purpose of mixing, that the convective eddies travel up 
to a distance which is some fractions of $H_p$ away from the formal
convective border. The numerical treatment is unchanged, the only
difference being the larger extension of the fully homogenized
region.
  
We thus have $l_{OV}=\alpha H_p$. According to 
Maeder \& Meynet (1991), a value of $\alpha=0.25$ is needed to 
fit the main sequences of young clusters; for older clusters the 
observed data are hardly fit with values of $\alpha$ exceeding 
0.2 (Stothers \& Chin 1992).

Within the diffusive framework it is necessary to specify the way 
with which convective velocities decay outside the convective 
boundaries (Deng at al. 1996a,b; Herwig et al. 1997; 
Ventura et al. 1998). 
In agreement with Xiong (1985), Grossman (1996)  and on the basis 
of numerical simulations by Freytag et al. (1996), we assume that 
convective velocities decay exponentially out of the formal 
convective boundary as:

\begin{equation}
$$
u=u_b exp \pm \left( {1\over \zeta f_{thick}}ln{P\over P_b}\right) 
$$
\end{equation}

\noindent
where $u_b$ and $P_b$ are, respectively, turbulent velocity and 
pressure at the convective boundary, P is the local pressure, $\zeta$ 
a free parameter connected with the e-folding distance of the decay, 
and $f_{thick}$ is the thickness of the convective regions in 
fractions of $H_p$. The chemical abundances are still calculated
by eq.(\ref{diffeq}), which is applied in a slightly more extended
region with respect to the borders fixed by the Schwarzschild criterion.

$\alpha$ and $\zeta$ are the two parameters connected with the 
extra-mixing in both schemes, though their physical meaning is 
completely different. In terms of duration of the main sequence 
phase for intermediate mass stars, we may say that $\zeta=0.02$
leads to results similar to $\alpha=0.18$, and $\zeta=0.03$ 
to $\alpha\sim 0.25$.

\subsection{Chemical composition}
We calculated several sets of stellar tracks by varying the 
scheme adopted for chemical mixing (instantaneous or diffusive), 
the convective model (FST and MLT), the parameter for the 
exponential decay of velocities ($\zeta=0.02$ and $\zeta=0.03$) 
and the rate of the reaction $^{12}C+\alpha \rightarrow ^{16}O$. 
Both for the instantaneous and the diffusive case a set of 
evolutionary sequences where the extension of the mixed region
is tightly fixed according to the Schwarzschild criterion was 
calculated. The adopted chemical composition for all the tracks 
discussed here is $Z=0.008$ and $Y=0.25$ (Brocato et al. 2003; 
Barmina et al. 2002).

\begin{figure}
\includegraphics[width=8cm]{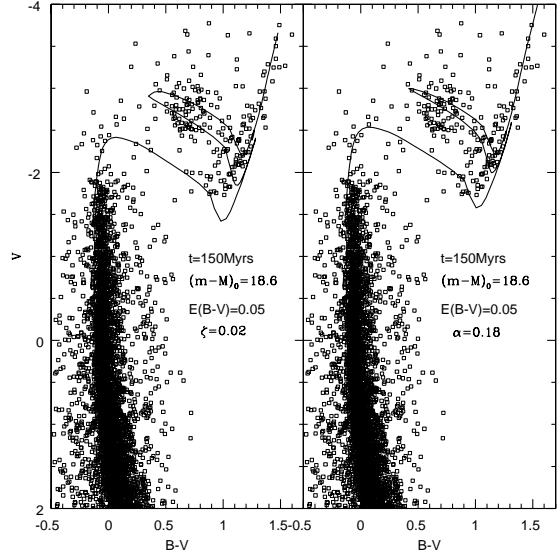}
\caption{The theoretical isochrones superimposed on the observational 
data for LMC cluster NGC1866. {\bf Left}: isochrones calculated 
with diffusive mixing and $\zeta=0.02$. {\bf Right}: instantaneous 
mixing, $\alpha=0.18$.}
         \label{CMdiagram}%
\end{figure}

\section{Some evolutionary remarks}
For all the sets of tracks considered we calculated theoretical 
isochrones for ages in the range $100 - 200$Myr following the 
scheme suggested by Pols et al.(1998).
The conversions from the theoretical to the observational plane 
have been performed by means of the transformations by Castelli 
et al.(1997). We first tried to select for each set of tracks 
the most appropriate parameters (age, distance modulus, reddening) 
allowing a decent fit of the observed CM diagram. First, we checked 
the possibility of discriminating between the two schemes for mixing,
looking for differences in the quality of fitting obtained with 
both prescriptions.
 
Fig.~\ref{CMdiagram} shows the comparison between the observed 
CM diagram (Testa et al. 1999) with the theoretical isochrones 
obtained by assuming a diffusive scheme for mixing with $\zeta=0.02$ 
({\it diff02} models), or an instantaneous prescription with an 
overshooting distance of $l_{OV}=0.18H_p$ ({\it ist18} models). 
The convective model adopted is FST, and the nuclear network is 
from Caughlan \& Fowler (1988). 

In both cases we see that a decent fit of both the extension of 
the main sequence and the luminosity and extension of the clump 
is obtained with an age of 150Myr, a distance modulus of 
$(m-M)_0=18.6$ and a reddening $E(B-V)=0.05$. 

\begin{figure}
\includegraphics[width=8cm]{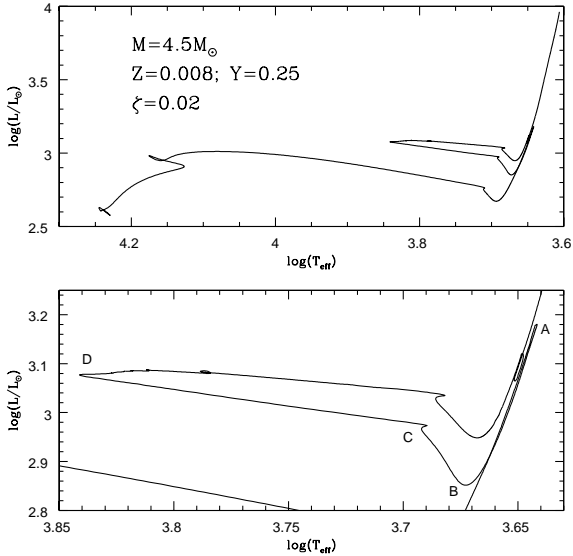}
\caption{The evolution on the theoretical HR diagram of a model
 of $M=4.5M_{\odot}$ calculated with diffusive mixing ($\zeta=0.02$).}
         \label{HR4.5}%
\end{figure}

This preliminary approach seems to rule out the possibility of 
selecting different mixing modalities simply on the basis of 
CM fitting. A deeper analysis is needed.
Before entering into the details of this comparison, we 
briefly examine the physical properties of the stars which 
are supposed to populate both the turn-off and the clump region 
within the context of our simulations. This will help us to 
better understand the results we obtain.
  
The typical mass belonging to the clump of fig.~\ref{CMdiagram}
is $4.5M_{\odot}$. The track on the HR diagram is shown in 
fig.~\ref{HR4.5}; the bottom panel shows in more details the 
clump region. In the early phases of hydrogen-burning the star 
develops a central convective region of $\sim 1.1M_{\odot}$ 
with a radial extension of $\sim 0.4R_{\odot}$. As CNO burning 
proceeds, the central regions contract and heat, the convective 
core shrinks in mass (it drops to $\sim 0.5M_{\odot}$ when the 
abundance of hydrogen is 0.25 in the centre) while 
its radius decreases slowly. Consequently, the pressure scale 
height at the border of the central convective region diminishes, 
and the extension of the extra-mixing region, directly connected 
to $H_p$, is also reduced. The whole phase of H-burning lasts 
$\sim 130$Myr, after which the star expands. Due to the lower 
and lower effective temperatures convection penetrates 
deeply towards internal regions which were previously touched 
by nuclear burning, in what is commonly known as the first 
dredge-up. The base of the external envelope reaches a layer 
which is $\sim 1M_{\odot}$ in mass away from the centre, while
the stellar luminosity, due essentially to CNO burning in a 
shell, rises up to $log(L/L_{\odot})\sim 3.2$, until 
$3\alpha$ reactions start in the centre of the 
star (point A in the bottom panel of fig.~\ref{HR4.5}).
At the ignition of helium in the centre the
energy release is in excess with respect to the
rate of energy which can be carried outwards. 
This triggers a rapid expansion of the central core,
which has the effect of diminishing the strength of the 
CNO shell burning, so that the luminosity decreases. The 
stellar layers from the shell to the surface contract, 
convection recedes, the track moves downwards 
in the HR diagram until the CNO shell, after $\sim 10^7$yrs 
since the beginning of helium burning, meets the chemical 
discontinuity left behind by the first dredge-up 
(point B).  This determines a rapid increase of the CNO (and global) 
luminosity, so that the star continues to burn helium in the red 
part of the HR diagram until surface convection disappears
(point C). Then a readjustment of the structure causes a rapid 
contraction, and the tracks moves to the blue part of the diagram 
until the CNO shell evolves through the discontinuity
(point D). The star remains in the blue part of the clump 
region until the helium abundance in the central regions 
drops to $Y \sim 0.2$, after which the rate of nuclear 
energy release drops and the core starts contracting; the
layers from the CNO shell to the surface expand, and the 
tracks moves to the red. In the {\it diff02} model the track 
stays in the bluest region of the clump for $\sim 7$Myr. 
The whole phase of helium burning lasts approximately 
$\sim 20$Myr.

\begin{figure}
\includegraphics[width=8cm]{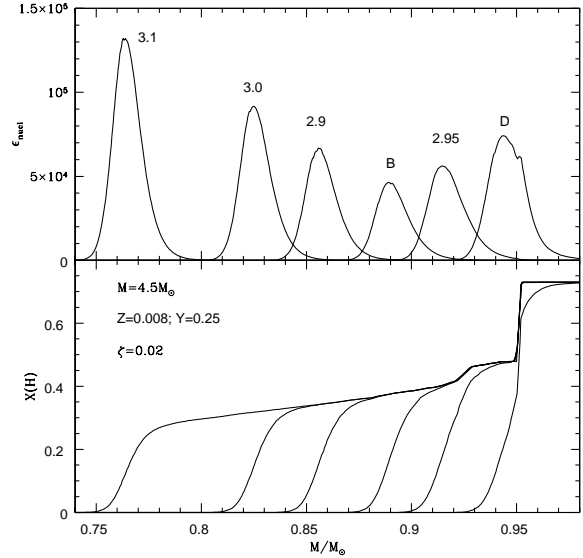}
\caption{The evolution with time of the internal structure of the
 same model of fig.~\ref{HR4.5} in terms of the coefficient of
 generation of nuclear energy (top) and the hydrogen profile (bottom)
 The numbers in the upper panel indicate value of the corresponding
 stellar luminosity, while characters B and D corresponds to the
 B and D evolutionary phases evidentiated in fig.~\ref{HR4.5}.
 We note in the bottom panel the chemical discontinuity at 
 $M/M_{\odot} \sim 0.95$.}
 \label{CNOprof}%
\end{figure}

Fig.~\ref{CNOprof} shows the variation with time of the
coefficient of generation of nuclear energy (top panel)
and the hydrogen profile (bottom) at some stages intermediate 
between points B and D of fig.~\ref{HR4.5}. We note
a rapid decrease of the energy release until the outer
part of the shell reaches the chemical discontinuity
(evidentiated by the sudden increase of the hydrogen
abundance at $M/M_{\odot} \sim 0.95$) (point B); 
point D corresponds to the phase when the peak of the 
shell reaches the quoted discontinuity.

We stress here the different behavior of the convective 
central regions of the star with respect
to the preceding phases of hydrogen burning. 
During H-burning, the extension in mass of the core tends
to increase as nuclear burning proceeds. The second 
important difference is that the ignition
of $3\alpha$ reactions provokes a rapid expansion of the central regions,
so that while the temperature increases, pressure and 
density decrease: this has the effect of
increasing the value of $H_p$ at the border of this region, with the 
consequence that more material external to the formal convective region 
is carried inwards.

This picture outlines which are the main factors driving the evolution 
of the star during the He-burning phase: 
\begin{itemize}

\item{
The excursion of the track to the blue is connected to the disappearance of the 
convective envelope.}

\item{
The bluest point of the track is reached when the whole CNO shell has evolved through
the chemical discontinuity left behind by the first dredge-up: we may thus expect that
the extension and the morphology of the blue loop depend on the mass coordinate of the
innermost point reached by external convection during the dredge-up and on the 
chemical stratification left.}

\item{
The return of the track to the red is connected with the contraction of the internal region
just before helium exhaustion. The duration of the permanence of the star in the blue is
therefore directly dependent on the amount of helium still present in the core when the
star reaches the bluest point of the loop. The way with which central regions are refurbished
of helium are expected to play a fundamental role in determining the time spent in the blue.}

\end{itemize}  

\section{Instantaneous vs. diffusive mixing}
Fig.~\ref{CMdiagram} shows that a decent fit of the HR diagram can be easily obtained both with an
instantaneous and a diffusive approach to deal with burning of chemicals in the regions
unstable to convection, because no clear difference can be seen in the theoretical
isochrones built with the {\it diff02} and {\it ist18} models.

To be able to distinguish between the results of the two schemes we have to build a
synthetic CM diagram, and to compare it with the observed CM diagram by Testa et al. (1999).
Our main goal is to reproduce the distribution of stars in the clump, i.e. to obtain
the observed ratio of $B/R\sim 1$ between the blue and the red stars of this region.


To build the synthetics, we developed a simple algorithm which populates a
given isochrone with a random distribution of masses, chosen to be available
for the isochrone of that given age. For each extracted mass, values of LogL and LogTe,
as well as magnitudes and colours in the observational planes,
were obtained by means of linear interpolation from the points of the same
isochrone. This simple way of populating an isochrone gives us important
indications regarding the relative populations of the various evolutive phases,
allowing us to derive the expected B/R ratio for each of our tests.

This ratio was calculated by counting the number of stars within
appropriate boxes in the CM diagram chosen on the basis of the observational
distribution of He-burning stars in NGC1866. Specifically, we retained
as blue stars those whose $(B-V)$ and $V$ fall within the range, respectively,
of $0.5 < (B-V) < 0.9$ and $15.6 < V < 16.3$; we considered as red stars,
those with $ 1 < (B-V) < 1.5 $ and $ 15.5 < V < 17.1 $.
We deliberately chose not overlapping boxes in order to prevent small
fluctuations in colour to substantially affect our statistics.

Statistical fluctuations related to the random distribution of star masses are
taken in account in the following way: for each choice of parameters, we performed
ten simulations varying only the seeds for random number generation: after that,
the computation of B/R ratio is made on the averaged numbers of blue and red,
according to the definition given above.
 
Note that we did not include the age spread in the following simulations. However,
preliminary computations show that a reasonable age spread ($\sim 10\%$ of
the age of the cluster) cannot alter
significantly the derived B/R ratio (in agreement
with Barmina et al. 2002, see their Fig.15).

\begin{figure}
\includegraphics[width=8cm]{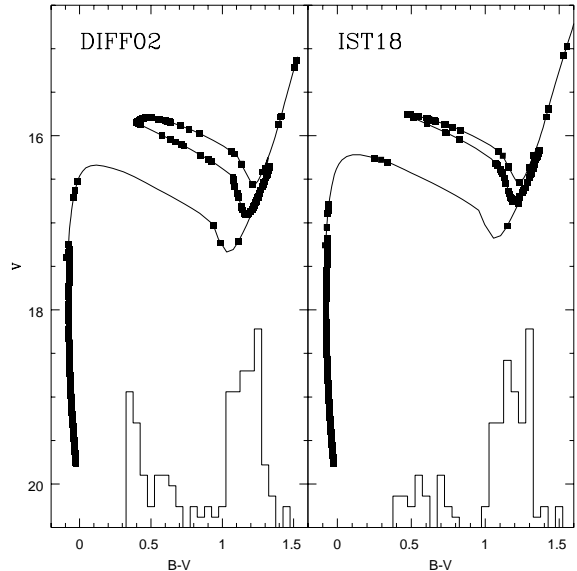}
\caption{Synthetic CMD computed by adopting diffusive stellar models 
         with $\zeta=0.02$ (left) and instantaneous models with
         $\alpha=0.18$ (right). In the lower part we show the hystohistogram
         of the distribution in colour of the He-burning stars; we
         note the presence of a blue population with $(B-V)$ in the
         range 0.4-0.5 in the {\it diff02} case.}
         \label{sint_1}%
\end{figure}

Fig.~\ref{sint_1} shows the numerical simulations obtained 
for the two sets of models: we can see that in the {\it diff02} 
case the blue part is much more populated, with a ratio 
$B/R\sim 60\%$, to be compared to the $B/R\sim 25\%$ value 
corresponding to the {\it ist18} models. This can be clearly
seen in the corresponding histogram shown in the lower part
of fig.~\ref{sint_1}. Such a steep dependence of the time spent
in the blue and red side of the CM diagram of He-burning 
intermediate mass stars on the treatment of mixing
within the convective core was already discussed in D'Antona 
(2002).

\begin{figure}
\includegraphics[width=8cm]{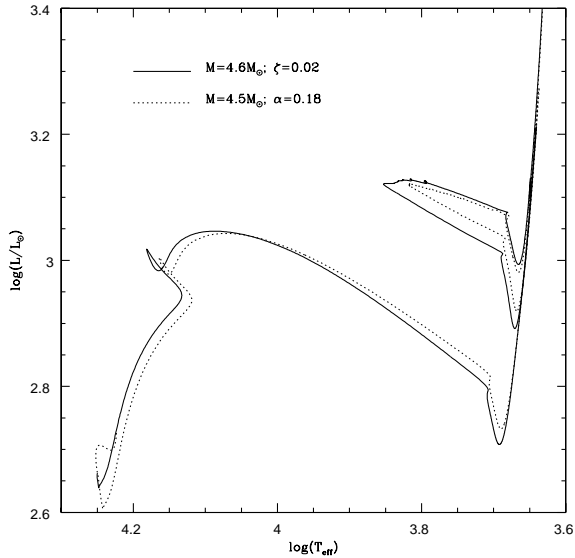}
\caption{The comparison of the tracks follower in the HR diagram
      by two models calculated with two different prescription for
      mixing. The selected masses populate the clump region according
      to our simulations.}
         \label{parmix1}%
\end{figure}

To understand the reason of this difference we compared 
carefully the evolutions of a
$4.5M_{\odot}$ model of the {\it diff02} 
set with a {\it ist18} model of $4.4M_{\odot}$, which 
are the typical masses staying in the clump at the age of $150$Myr for the two sets
of model discussed. The two tracks are very similar during the MS evolution 
(see fig.~\ref{parmix1}), and have approximately the same luminosity during the helium 
burning phase. The duration of the H-burning phase is the same ($\sim 130$Myr) 
for both models. 

\begin{figure}
\includegraphics[width=8cm]{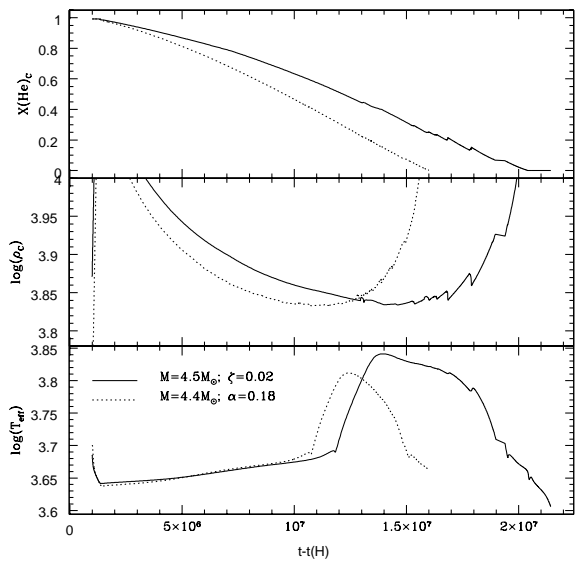}
\caption{The variation with time of some quantities related to the
         evolution of the same models shown in fig~\ref{parmix1}.
         In the abscissa we report the time counted from the hydrogen
         exhaustion in the stellar centre. {\bf Upper}: Central helium;
         {\bf Middle}: Central density. {\bf Lower}: Effective 
         temperature.}
         \label{parmix2}%
\end{figure}

A detailed comparison between the two evolutions can be seen in the three 
panels of fig.~\ref{parmix2}.
In the abscissa we report the time starting from the beginning of 
hydrogen exhaustion in the centre; to have an idea of the extension 
of the loop and the duration of the blue phase, we show in the lower 
panel the variation of the effective temperature. We find
in both cases a rapid increase of $T_{eff}$ with time in conjunction 
with the disappearance of surface convection. 
The diffusive track stays in the blue longer ($t_{blue} \sim 7$Myr), 
while the {\it ist18} track, as soon as it reaches the bluest point, 
moves again to the red ($t_{blue} \sim 3$Myr).

The reason of such a different behavior can be detected in the 
first and second panel of the same figure. In the upper panel 
we see that in the {\it ist18} model helium is burnt faster, 
so that when the bluest point is reached the central helium 
abundance has dropped to $\sim 0.2$;
with these low helium abundances the two nuclear reactions involved 
($3\alpha$ and $^{12}C+\alpha \rightarrow ^{16}O$) cannot supply 
the whole energy demand, so that the contraction of the core, 
triggering the expansion of the external 
layers and the return of the track to the red, starts soon. 
In the diffusive model the helium burning is slower, 
so that at the bluest point there is still 
$40\%$ of helium in the core, thus delaying the beginning of 
core contraction. 
This allows the track to stay in the blue region for longer, and the 
relative duration of this phase is approximately doubled.

The results are in full agreement with the detailed 
comparison of the H and He-burning times obtained with 
the two schemes for mixing for the most massive intermediate
mass stars presented in Salasnich et al.(1999) (see their fig.5). 

These results indicate that the two schemes for mixing can lead to
deeply different results, particularly in terms of the relative duration
of the various evolutionary phases. The diffusive scheme allows an 
exponential decay of velocities beyond the border of convection,
with the consequence that the zone which is mixed is larger:
yet, we must consider that the time-scales for mixing 
in the most external layers are extremely large,
so that the refurbishment of fresh helium into the hottest central
layers is slow. On the contrary, in the instantaneous case a
smaller region is involved in mixing, but the efficiency of convection 
in this latter case is assumed to be so high to homogenize it
completely: the process of carrying helium to the centre is thus 
more efficient, the rates of both $3\alpha$ 
and $^{12}C+\alpha \rightarrow ^{16}O$
reactions are increased, and the whole process of helium burning is faster.

It is not surprising that the differences between the results obtained with the
two schemes for mixing are more evident during the He-burning phase,
since in this case the size of the convective core tends to increase;
during the H-burning phase the shrinking of the convective core, precedently
discussed, makes the size of the extra-mixing regions to be progressively
decreased, so that the differences between the two treatments are 
smaller.

The comparison between the results obtained with different schemes
for mixing leads to the conclusion that intermediate mass models
computed with the diffusive approach, provided the luminosity of
turn-off and of the clump are the same, consume helium more slowly 
than the corresponding models computed with the instantaneous 
mixing approximation, with a consequent longer duration of the blue 
phase.

Applied to the study of NGC1866, the usage of the diffusive approach
allows to shorten the gap between theory and observations, rising the 
predicted $B/R$ ratio to $\sim 60\%$ (to be compared to the expected
value $B/R \sim 25\%$ on the basis of the {\it ist18} models).

\subsection{The no extra-mixing case}
The same analysis to test the influence of the time-dependent
algorithm for mixing was applied to two sets of 
tracks calculated by assuming that
no sort of extra-mixing takes place out of the border fixed by the
Schwarzschild criterion ({\it diff00} and {\it ist00} models).
 
\begin{figure}
\includegraphics[width=8cm]{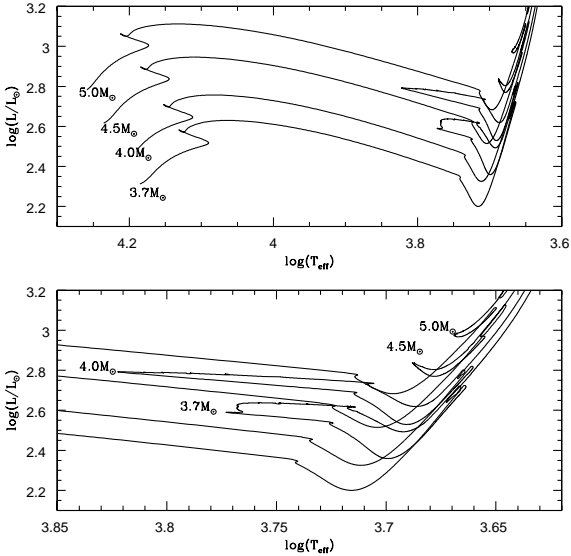}
\caption{Evolutive tracks of {\it ist00} models with masses in the range 
         3.5 - 5$M_{\odot}$ in the theoretical plane. We note the very
         narrow extension of the loop for the most massive models shown.}
         \label{ist00tracks}%
\end{figure}

Fig.~\ref{ist00tracks} shows the tracks followed in the HR diagram
by models with masses in the range 3.5 - 5$M_{\odot}$, calculated
with the instantaneous mixing approximation. We note that the bluest
point of the track gets hotter as the stellar mass increases, but
for models more massive than $\sim 4.3M_{\odot}$ the track substantially 
stays in the red during the whole He-burning phase; this behavior
is found for all models less massive than $\sim 6.5 M_{\odot}$.
These results are in full agreement with the most recent findings
concerning the evolution of intermediate mass stars of this
metallicity by Pietrinferni et al. (2004).

This can be explained on the basis of the fact
that in the more massive models helium consumption if faster (due to
the higher temperatures) and convection is always well
extended: when the abundance of helium drops to $Y \sim 0.2$,
and the core contraction begins, the surface convection is still alive, 
so that the star never begins the rapid excursion to the blue which is 
typical of models with radiative envelopes. Therefore, we
find no blue clump stars for ages $t \leq 140$Myr.

Disregarding the presence or not of the blue loop,
on the basis of the clump luminosity we obtain an 
age for NGC1866 of $\sim 100$Myr with a distance modulus 
of $(m-M)_0\sim 18.5$, in agreement with Testa et al. (1999); 
for a lower distance modulus ($(m-M)_0\sim 18.25$) we obtain 
a decent fit of the clump luminosity for an age of $\sim 150$Myr
(in agreement with Brocato et al. 2003).

The lack of stars burning helium in the blue 
seems to rule out ages $\leq 140$Myr, and in any case does not 
allow us to perform the usual analysis based on the $B/R$ ratio.
For an age $t=150$Myr we cannot
fully reproduce the extension of the clump both in temperature 
and luminosity; in any case, the percentage of blue stars 
compared to the red would be low ($B/R \sim 25\%$) compared 
to the observed value of $\sim 1$. This can be understood on
the basis of the helium still present within the core when
surface convection is extinguished. The lack of any extra-mixing
out of the formal convective border triggers a rapid consumption 
of helium in the central region, so that when the excursion
towards the blue begins the abundance of helium dropped to
$Y \sim .25-.3$: soon after the contraction of the central
layers forces a quick return of the track to the red. 

It's interesting to note that the B/R ratio
expected on the basis of the instantaneous models
calculated with no extra-mixing is the same
as in the {\it ist18} models, precedently discussed.
Therefore, within the framework of instantaneous models,
it is not possible to derive informations
concerning the extension of the overshooting region
on the basis of this kind of analysis only.

If we compare the results obtained with the
two schemes for mixing ({\it diff00} and {\it ist00} models), 
we note that in these models calculated
by assuming a complete absence of extra-mixing no meaningful
differences have been found, both in the main sequence and in the
helium burning phase. This can be explained on the basis of the 
fact that the time scale of convection is very short with respect
to those typical of the various reactions involved within the
formally unstable region. For example, within the $4.3M_{\odot}$ 
{\it ist00} model, which would populate the clump of NGC1866 
if no extra-mixing is considered, the time-scale of convection 
is always shorter than $\sim 1$ month for the whole duration 
of the He-burning phase.

\section{The role of $\zeta$}
The discussions of the previous sections outlined the difference between 
the roles of the parameters $\zeta$ and $\alpha$ in determining the 
amount of matter which is mixed and carried to the central regions 
during the major phases of central burning.

$\zeta$ is connected with the e-folding distance of the velocity decay 
from the border of the convective zone, but even in case of very large 
values of $\zeta$ material far from the border is hardly mixed to 
the centre, because the time-scale for mixing in that region would 
be comparable or even longer than the evolutive times. Hence, we 
expect that even large percentage variations of $\zeta$ 
don't have dramatic consequences on the evolution of the star, unlike 
the effect of great changes of $\alpha$.

\begin{figure}
\includegraphics[width=8cm]{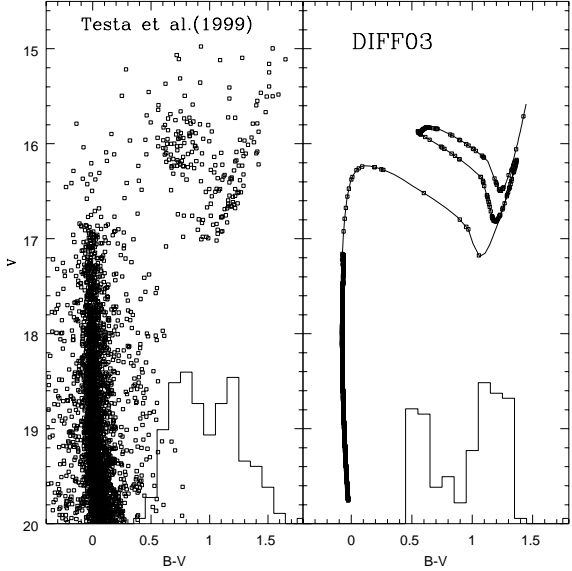}
\caption{Synthetic CMD computed by adopting diffusive stellar models 
         with $\zeta=0.03$ (right) compared with the observed CMD
         of NGC1866 by Testa et al. (1999). In the lower part of both
         panels we show the corresponding histogram related to the
         distribution in colour of the He-burning stars.}
         \label{sint_2}%
\end{figure}

A preliminary study regarding the effects of varying the e-folding 
distance of the exponential decay of velocities outside the border
of the convective core is found in Salasnich et al.(1999), but it
is mainly focused on the evolution 
of massive stars ($M\sim 20M_{\odot}$).
A straight detailed comparison between our parameter $\zeta$ and
their quantity $\alpha_1$ is made difficult by the different way
of simulating the exponential decay: we use pressure as independent 
variable, while they use the radius.

We increased the value of $\zeta$ up to 0.03, obtaining a nice fit of the 
CM diagram of NGC1866 with the same parameters holding for $\zeta=0.02$, 
but with a slightly larger age ($160$Myr). In this latter case the mass 
evolving in the clump is $M_{CLUMP}\sim 4.4M_{\odot}$. The numerical 
simulations run for this set of tracks show (see fig.~\ref{sint_2}) 
that the percentage of blue stars is increased up to $B/R\sim 75\%$, 
which is still lower than the observed value, but pointing in the right
direction.

\begin{figure}
\includegraphics[width=8cm]{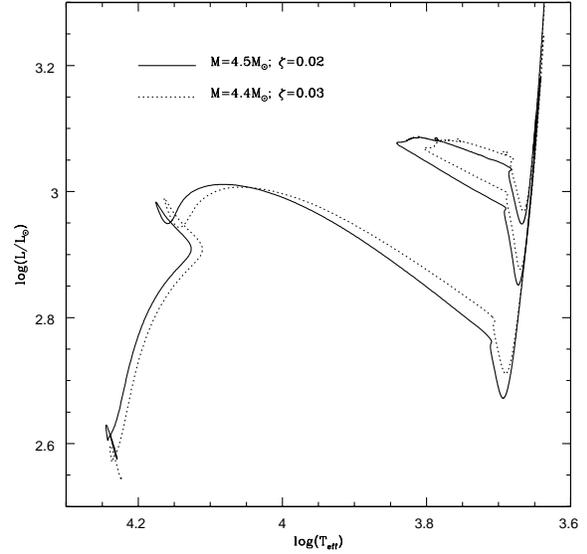}
\caption{Path followed by two models calculated with the diffusive
         scheme for mixing and different values of $\zeta$ which
         are currently populating the clump region according to our
         simulations.}
         \label{pareta1}%
\end{figure}

\begin{figure*}
\centering{
\includegraphics[width=8cm]{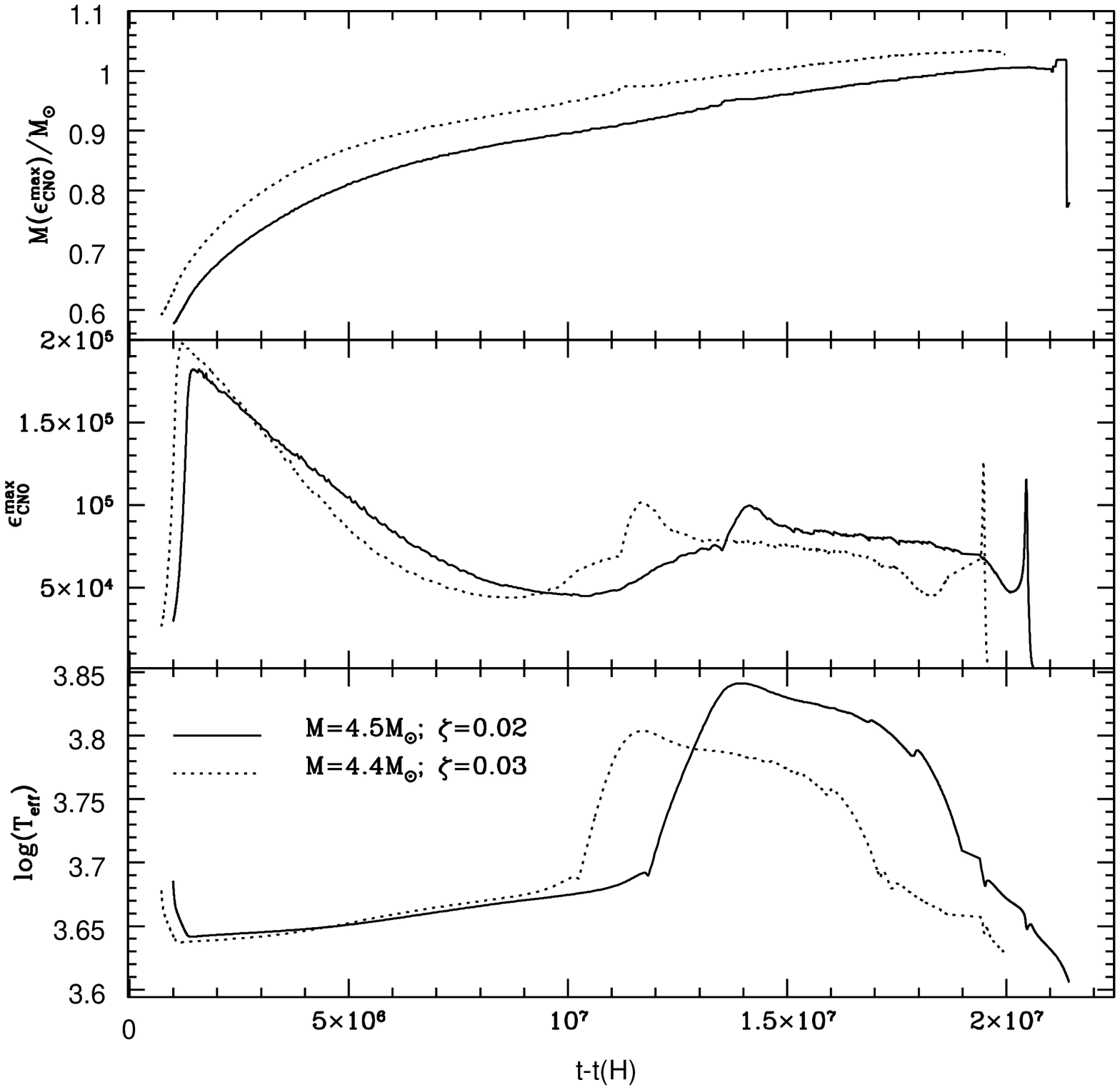}
\includegraphics[width=8cm]{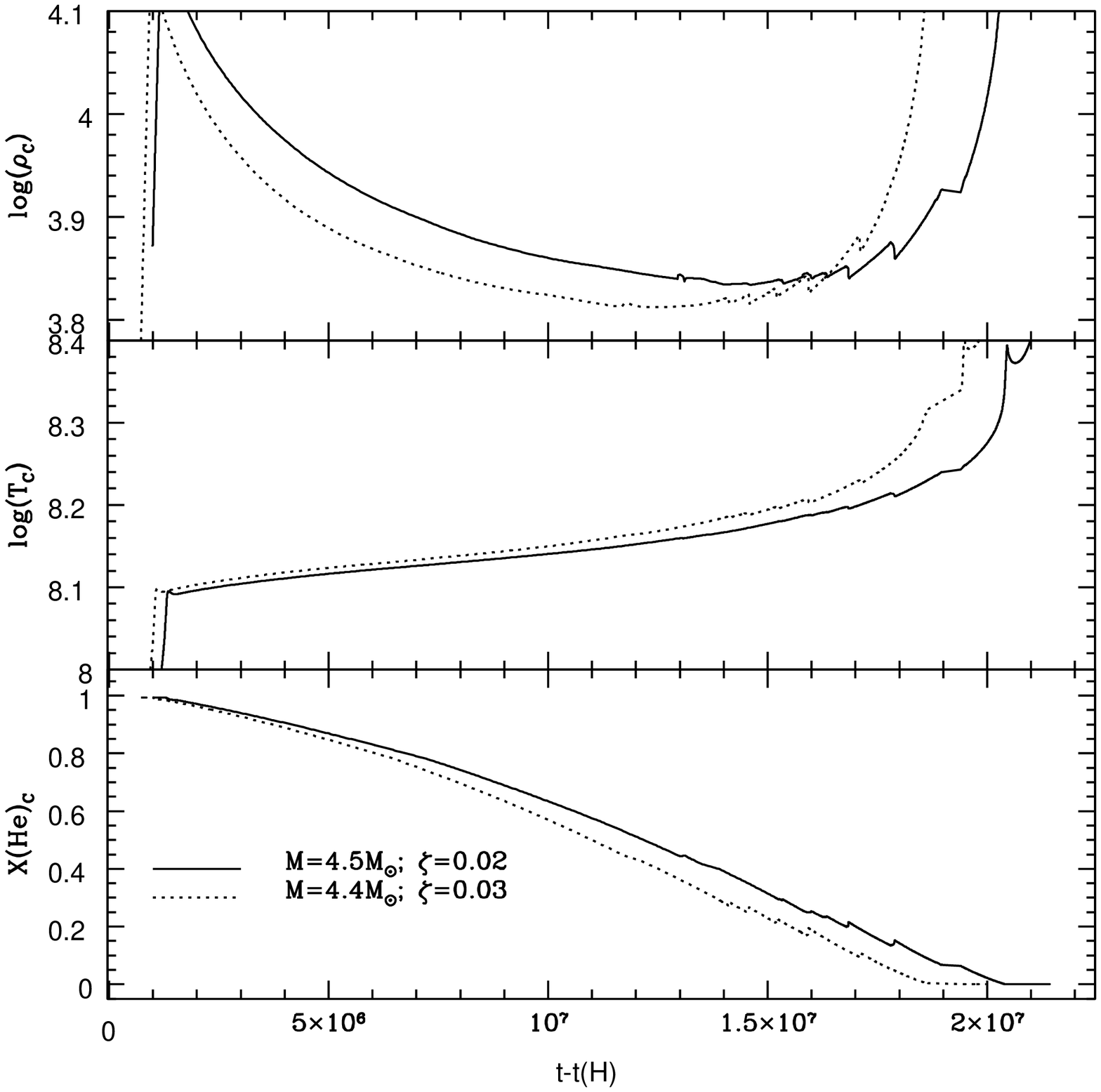}
           }
\caption{Variation with time of some quantities related to the same
         models shown in fig~\ref{pareta1}. In the abscissa we report
         the time counted from hydrogen exhaustion in the centre of the 
         star.}
   \label{pareta2}
\end{figure*}

The increase of the time spent in the blue for larger values of $\zeta$ 
can be understood again if we consider the fast consumption of helium 
within the central regions, and the quantity of helium 
still left when the track reaches the bluest point of the clump.

To understand the differences between the $\zeta=0.02$ and $\zeta=0.03$ sets 
we compare two models of masses $M=4.4M_{\odot}$  (case $\zeta=0.03$) 
and $M=4.5M_{\odot}$ ($\zeta=0.02$).  
Such a choice stems from the small difference found between the masses
populating the clump region in the best simulations obtained with
the two sets of models.

In fig.~\ref{pareta1} we compare the tracks on the HR diagram; 
we note that both the luminosity of the turn-off and of the bluest point 
of the clump are almost the same.
In the 6 panels of fig.~\ref{pareta2} we show in details the differences 
in terms of various chemical and physical quantities during the 
He-burning phase.

In the left panels we report the evolution with time of the CNO burning shell 
in terms of the mass coordinate of the layer corresponding to the maximum 
release of CNO nuclear output (left - upper panel) and of the maximum 
value of the coefficient for the generation of CNO nuclear energy 
(left - middle panel). The right panels shows the evolution of the central 
physical-chemical conditions, in terms of density, temperature, 
and helium abundance.

In the $\zeta=0.03$ model the H-exhausted region is larger, and the
CNO shell is more external (in mass); the difference is of the order
of $\Delta m \sim 0.04M_{\odot}$. The CNO burning shell in the $\zeta=0.03$
model, before the ignition of $3\alpha$ process, is consequently hotter 
and more efficient, so that the energy release is larger, as we can see in
the left - middle panel. The faster drop of CNO energy release 
 which can be seen in the $\zeta=0.03$ model, is related to the conditions
in the core. The right-middle panel of fig.~\ref{pareta2} 
evidentiates that the central 
regions of the $\zeta=0.03$ model are slightly hotter, 
so that  helium is consumed faster (right-lower panel);
this, in turn, triggers a more rapid expansion of the central regions, 
which can be detected in the steeper drop of central density in the 
right-upper panel,  and a more rapid decrease of the nuclear energetic
output due to CNO burning shell (see the left-middle panel).
 
The evolution of the $\zeta=0.03$ model, for what concerns the 
conditions of the external layers of the structure, is faster, so 
that the overall contraction leading the track to the blue in the
HR diagram is anticipated by $\sim 2$Myr with respect to the
$\zeta=0.02$ case.

The same difference in the velocity of the evolution is not
found in the central regions, because the larger
temperatures within the interior of the $\zeta=0.03$ model
(which would lead to a more efficient nuclear burning  and thus 
to a faster helium consumption) are partly balanced by the 
larger value of $\zeta$, which allows to refurnish the core 
with more helium from the regions
away from the formal border of the convective core to be transported 
in the hot central regions. The earlier excursion 
to the blue of the $\zeta=0.03$ track is not
followed by a quicker return to the red (see the left - lower panel 
of fig.~\ref{pareta2}) so the percentage of time spent in the
blue increases ($\sim 70\%$, to be compared to
$\sim 50\%$ for $\zeta=0.02$).

We also note that the bluest point of the $\zeta=0.03$ model
nicely fits the location on the HR diagram of the  
bluest Cepheid EV Sct found in the open cluster NGC 6664
(Schmidt 1984).

These results, together with the 
conclusions drawn at the end of the previous section,
show that the diffusive model with $\zeta=0.03$ leads
to a better agreement between the theoretical framework
and the observational scenario. Thus, we decided to adopt
this set of models to perform the further analysis to test
the influence of the convective model and of the relevant
reaction rates.

\section{Can the CM diagram of NGC1866 help in discriminating 
         between convective models?}

The determination of the temperature gradient (or, equivalently, of
$\nabla={dlogT\over dlogP}$) within convective regions is
one of the most disputed issues in modern stellar astrophysics
(Ulrich 1976; Grossman 1996; Grossman \& Narayan 1993;
Xiong et al. 1997; Canuto 1997; Canuto \& Dubovikov 1998).

These difficulties stem from the absence of an adequate solution of
the Navier -- Stokes equations, so that the physical description of
the instability regions still rely on local approximations, in which 
$\nabla$ is calculated on the basis of the values which some physical
quantities assume in the same point where we want to know $\nabla$.

This approximation neglects one of the most important features of
turbulence, i.e. non locality, but still it is the only opportunity
allowing the evaluation of $\nabla$.

Canuto \& Mazzitelli (1991) constructed a local model for
turbulent convection (FST model) which has been successfully employed
in several astrophysical studies, like helioseismology
(Basu \& Antia 1994), red giant evolution (Stothers \& Chin
1995; 1997), AGB evolution of intermediate mass stars
(D'Antona \& Mazzitelli 1996; Mazzitelli et al. 1999).
Within this context we test the possibility of selecting 
between the two local convective models currently available,
i.e. the Mixing Lenght Theory (MLT) and the FST model, 
on the basis of the comparison between the theoretical predictions 
and the observed distribution of NGC1866 stars in the CM diagram.

We therefore calculated a set of MLT models with a diffusive scheme for 
mixing; the parameter for the exponential decay of convective velocities
was set to $\zeta=0.03$, on the basis of the discussion presented in the 
previous section.

\begin{figure}
\includegraphics[width=8cm]{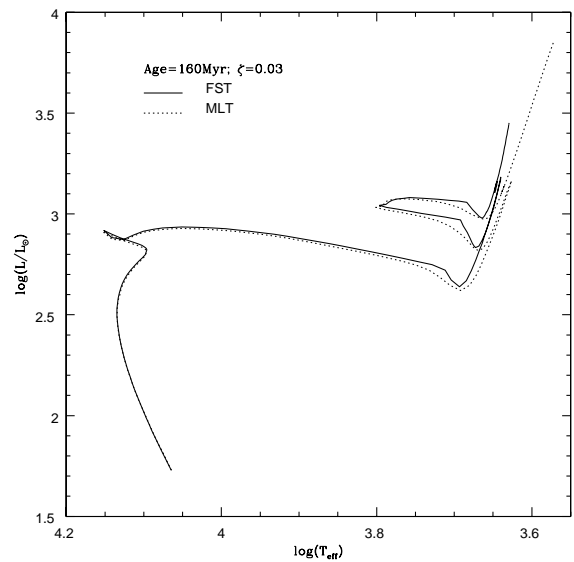}
\caption{The comparison between two isochrones of the same age corresponding
         to models calculated with two different convective schemes.
         We note the different slope of the giant branches, the MLT
         isochrone becoming cooler and cooler for lower $T_{eff}$'s.}
         \label{par_conv_1}%
\end{figure}

In fig.~\ref{par_conv_1} we compare the isochrone 
giving the best fit with the $\zeta=0.03$ FST models 
with the corresponding isochrone built by using the MLT models. 
The two lines are almost perfectly superimposed
during H-burning, while we note a different slope of the 
giant branch, the MLT line becoming progressively cooler for 
lower effective temperatures. The luminosity and the extension 
of the clump are very similar. 

The numerical simulations (not shown here) 
show no meaningful differences in the distribution of 
stars in the clump, particularly in terms of the $B/R$ ratio; 
the only difference between the two sets of models is in the colour 
of the giant branch, which is redder in the MLT models.
The sensitivity of the colors of giants to the convective
model adopted has been explored
by Stothers \& Chin (1997) (see also the discussion in
Asida 2000), who found that a further tuning of the
mixing length parameter with mass and effective temperature
is needed to fit the observed slopes of red giant branches.
We recall that in the present MLT computations the
solar calibrated mixing length parameter was used 
for all the evolutions.

On the theoretical point of view these results are indeed 
not surprising, if we consider the physical
contexts where we expect the convective model used to have an impact 
on the results obtained, in terms of temperature stratifications.
Actually, inside convective regions where the efficiency of convection
is very large (i.e. almost all the energy flux is carried by convection),
the gradient sticks to the adiabatic value, and this is almost
independent of the description of convection adopted.

In the case of NGC1866, if we consider the evolution of the masses 
currently in the clump according to our simulations 
(i.e. $M\sim 4.5M_{\odot}$), we can select three different cases 
where the star develops wide convective regions:

\begin{itemize}

\item{The convective core during H-burning.}

\item{The convective central region which is formed along
with the ignition of $3\alpha$ processes.}

\item{The convective envelope which forms following the
  expansion of the external layers after hydrogen exhaustion.}

\end{itemize}

\noindent

In the first case (core H-burning), the overadiabaticity 
$(\nabla - \nabla_{ad})$ is
on the average $\sim 10^{-7}$ within the whole convective core in
both models; the only difference can be noticed just in the very
proximities of the border, where the FST $(\nabla - \nabla_{ad})$ 
is seen to rise up to $10^{-5}$, while in the MLT case we find 
$(\nabla - \nabla_{ad}) \sim 10^{-8}$. 
This different behavior can be understood on the
basis of the different mixing scales adopted in the two cases (Canuto \& 
Mazzitelli 1991). In the
FST model the choice $\Lambda = z$, the distance from the border,
leads to a lower convective efficiency, while the MLT, adopting 
a typical scale proportional to $H_P$, simulates a much larger 
efficiency of convection in these regions.
Even taking into account these differences, 
we still have a very small degree of overadiabaticity in a region 
where the pressure drop is approximately 2 dex ($log(P)\sim 17$ 
in the centre, and $log(P)\sim 15$ at the border), so that in 
both cases the temperature gradient sticks to the adiabatic value. 
This can explain the similarity of the
two isochrones in the MS region of the cluster.

In the second case (core He-burning) we expect the 
differences between the results 
provided by the two prescriptions for convection to be even less
than during H-burning, because of the larger densities in the 
He-burning core, leading to a larger efficiency of convection.
Indeed, we find in this case that along the whole convective core
$(\nabla - \nabla_{ad}) \sim 10^{-8}$, with a maximum value of
$(\nabla - \nabla_{ad}) \sim 10^{-7}$ attained by the FST model in
the regions next to the border. We may conclude that even during
He-burning the temperature stratifications are very close, so that
the differences in terms of duration of the whole He-burning 
phase are completely negligible.

The third point quoted above (convective envelope) deserves 
particular attention, because the outermost 
layers of the star, being cooler and less dense than the 
interiors, present values of $\nabla$ which largely deviates 
from the adiabatic values. 

Within the convective envelope of the two models precedently
discussed we note a mean value of overadiabaticity which
is $(\nabla - \nabla_{ad}) \sim 3\times 10^{-5}$ in the FST
case, and $(\nabla - \nabla_{ad}) \sim 3\times 10^{-4}$ in the
MLT model: this difference is due to the larger efficiency
of convection of the FST model, which leads to larger convective
fluxes for a given value of overadiabaticity 
(Canuto \& Mazzitelli 1991).  
These values are considerable
higher than those found within the central cores; this, together 
with the severe drop of pressure from the base of the external 
zone up to the surface ($\sim 5-6$ dex) makes the differences 
in terms of temperature stratification more evident. 

In the very proximities of the surface
the FST model has an overadiabaticity peak which is
higher and narrower than the MLT one: this is again due to the 
MLT choice of the scale of mixing, which acts to simulate 
a convective efficiency which keeps sufficiently large even 
in regions where in reality a very low percentage of energy 
is carried by convection. Yet, these differences involve only 
the outermost layers, therefore do not have any incidence on 
the results obtained.

We may conclude that in the MLT case the slope of the temperature
drop from the base of convection to the surface is higher, so 
that the effective temperature is lower; this becomes more evident 
for deeper convective envelopes, so we may understand the 
different slope of the giant branches which can be seen in 
fig.~\ref{par_conv_1}.

On making the comparison with the observed CM diagram of 
NGC1866, we don't find any appreciable difference in terms of 
the $B/R$ ratio, for two main reasons:

\begin{itemize}

\item{During the evolution in the blue the only convective zone
within the star is the He-burning core, for which we already stressed
that the gradient is almost adiabatic: the evolution of the star
in terms of evolutive time and colour is therefore independent of
the convective model.}

\item{In the red region of the clump, we saw that the convective
model directly influences the values of $T_{eff}$, but not of the 
thermodynamical quantities of the CNO burning shell; the time spent
in the red is therefore not influenced by the convective model.}

\end{itemize}

This comparison between the observational scenario and the theoretical 
predictions for two convective models can only confirm the problems
of fitting the colours of giants connected with the MLT model, but,
since the influence upon the B/R ratio is very poor, it is not
possible within this context to discriminate between convective models.

\section{The reaction $^{12}C+\alpha \rightarrow ^{16}O$.}
The evolution of a star in the early phases of He-burning is dominated
by the $3\alpha$ reactions, but as helium is consumed in the central regions
the reaction $^{12}C+\alpha \rightarrow ^{16}O$ becomes more and more important,
eventually determining the duration of the last phases of helium burning and
the residual chemical abundances left in the central regions of the star
after helium exhaustion.

\begin{figure}
\includegraphics[width=8cm]{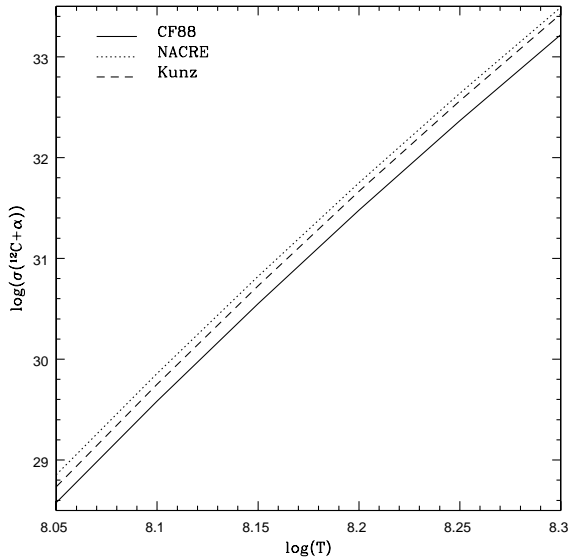}
\caption{The variation with temperature of the cross
section (expressed in cm$^3$s$^{-1}$mole$^{-1}$ units)
of the reaction $^{12}C+\alpha \rightarrow ^{16}O$
according to the prescriptions given in Caughlan \&
Fowler (1988, CF88), Angulo et al. (1999, NACRE), and 
Kunz et al. (2002). }
   \label{sigmac12}
\end{figure}

\begin{figure}
\includegraphics[width=8cm]{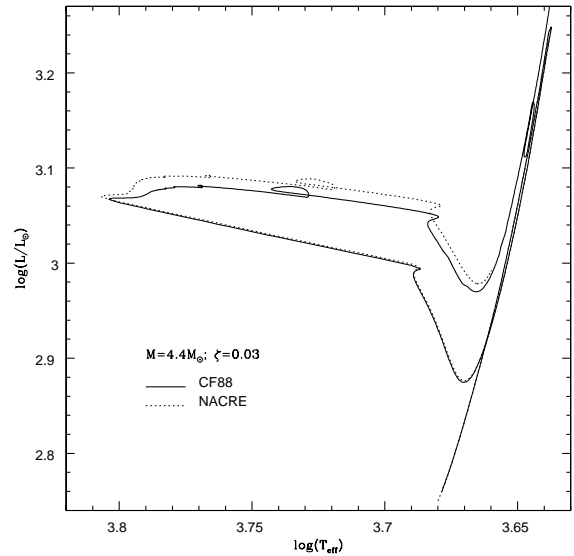}
\caption{The path followed in the HR diagram by three models of
mass $M=4.4M_{\odot}$ calculated with different cross-sections
for the nuclear reaction $^{12}C+\alpha \rightarrow ^{16}O$.
The solid track refer to the cross-section taken by Caughlan \&
Fowler (1988), the dotted line has been calculated with
the NACRE rate by Angulo et al. (1999), and the dashed line
with the rate given by Kunz et al. (2002).}
   \label{par_c12_1}
\end{figure}

\begin{figure*}
\centering{
\includegraphics[width=8cm]{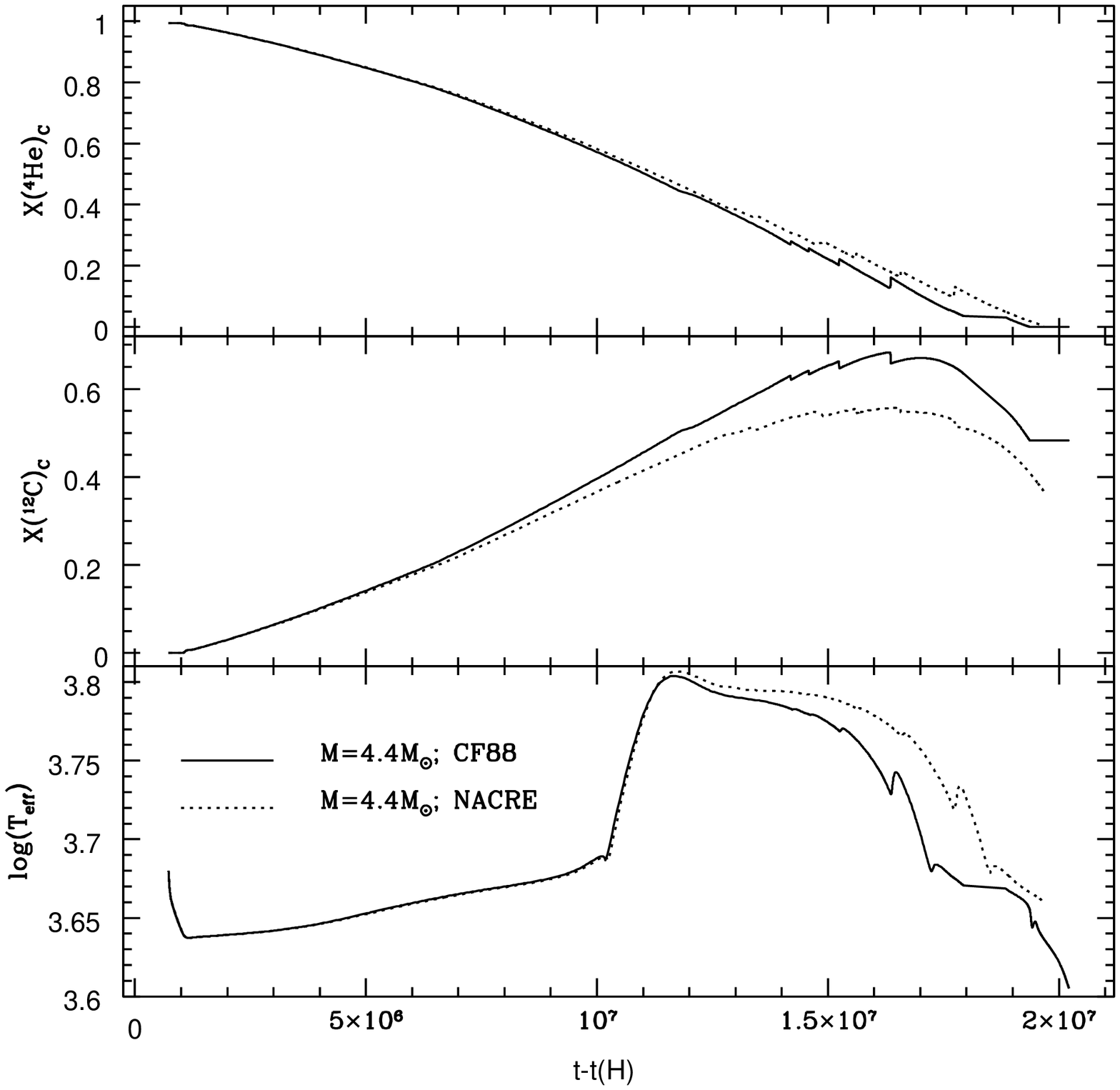}
\includegraphics[width=8cm]{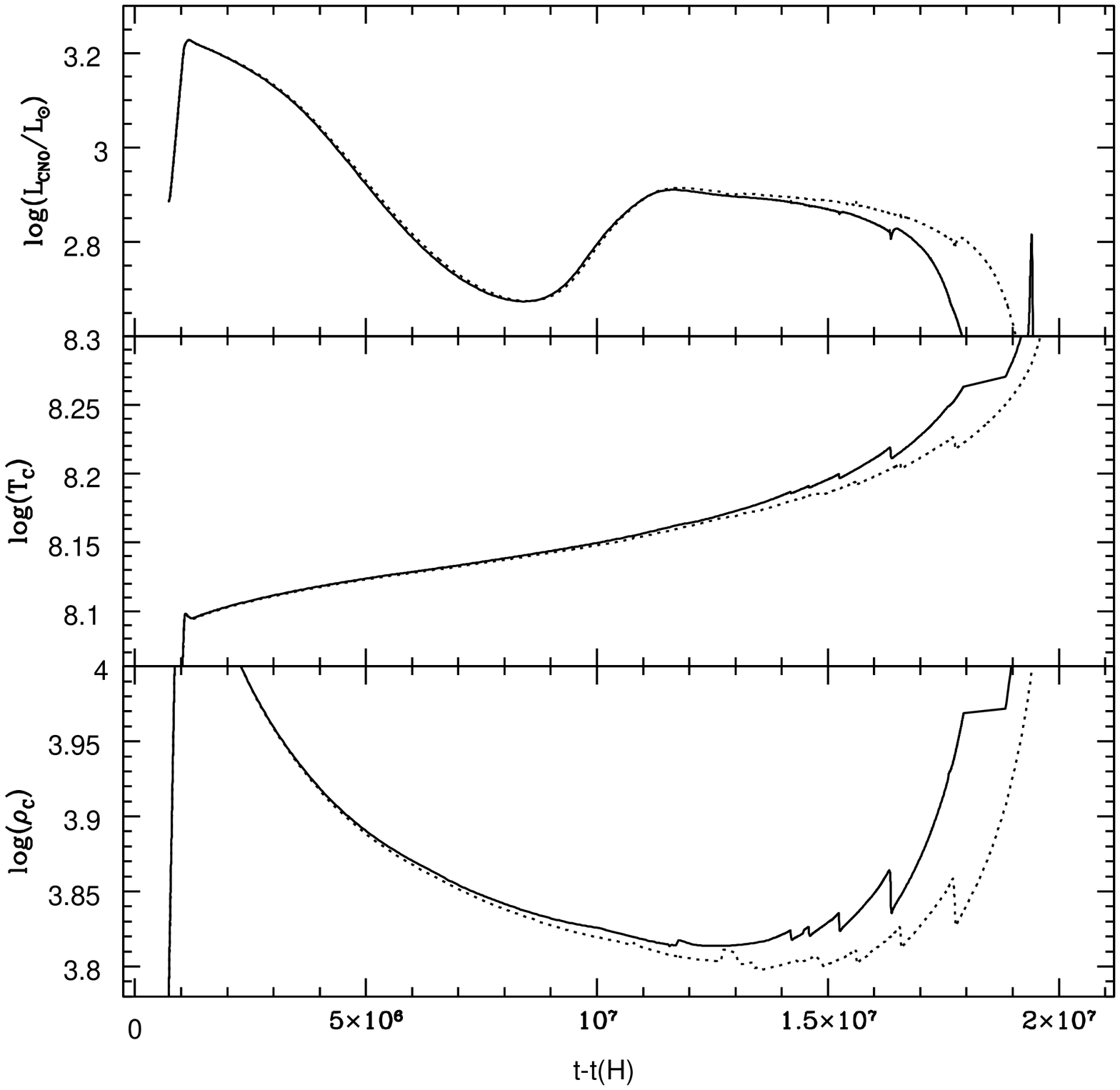}
           }
\caption{The variation with time of various chemical and physical
   quantities related to the evolution of the same models shown
   in fig.~\ref{par_c12_1}.}
   \label{par_c12_2}
\end{figure*}

The cross-section $\sigma_{^{12}C+\alpha}$ is uncertain by 
at least a factor 2, so we want to check the sensitivity of the expected 
distribution of He-burning stars in the CM diagram which may be due 
to changes in the adopted relationship $\sigma_{^{12}C+\alpha}(T)$.

The influence of the quoted reaction on the times 
spent by intermediate mass stars in the helium burning 
phase, and particularly on the blue region
of the clump, has been explored by Imbriani et al.(2001), who
compared the results obtained by adopted either 
the Caughlan \& Fowler (1988, hereinafter CF88) 
or the Caughlan et al. (1985) rates, 
this latter release providing approximately rates larger
by a factor of 2. Their main findings, essentially, were 
that the tracks in the
HR diagram are not dramatically altered by the choice 
of $\sigma_{^{12}C+\alpha}$,
while the relative duration of the blue phase during helium burning is 
increased by $\sim 10\%$. This result would thus increase the percentage of 
blue stars in the clump region, helping to reconciliate the observational
scenario for NGC1866 with theoretical predictions.

\begin{figure}
\includegraphics[width=8cm]{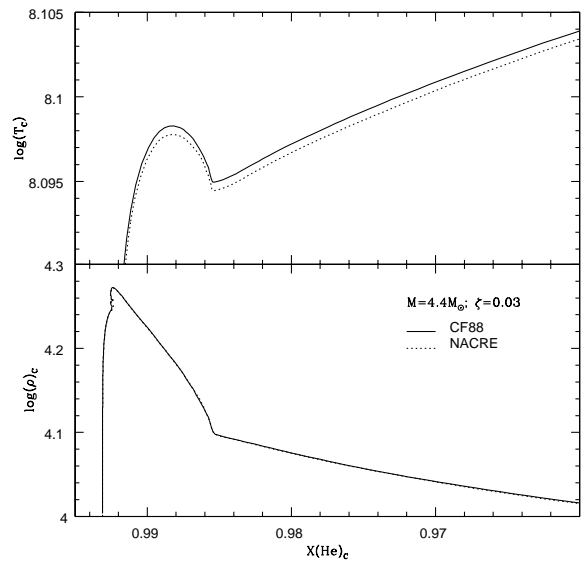}
\caption{Variation of the central thermodynamical conditions
 of two of the three models presented in fig.~\ref{par_c12_1}
 during the early phases of helium burning.} 
   \label{par_c12_3}
\end{figure}

Here we want to compare the results obtained by adopting 
the CF88, the NACRE (Angulo et al. 1999) and the Kunz et al.
(2002) rates for  the quoted reaction. We therefore calculated the same 
stellar models giving the best fit of the observational
scenario (i.e. diffusive models with $\zeta=0.03$) with the NACRE 
$\sigma_{^{12}C+\alpha}(T)$ relationship, which provides 
values larger by a factor of $\sim 2$ over the whole 
range of temperatures of interest here (Angulo et al. 1999),
and with the Kunz et al. (2002) prescription, which provides
cross-sections sligtly lower than NACRE, and more sensitive
to temperature changes in the range $10^8 < T(K) < 2\cdot 10^8$,
which are those of interest here (see fig.~\ref{sigmac12}).

Fig.~\ref{par_c12_1} shows the comparison between three
tracks of a model of mass $M=4.4M_{\odot}$, the typical 
mass populating the clump region (see Sect.5),
calculated with the three different reaction rates. We focus 
only on the clump region, since no differences are found 
during the precedent phases. We see that the morphology 
and the extension of the loops are very similar, with
the only difference that the NACRE model is slightly more 
luminous in the upper part, which corresponds to the phase 
of expansion of the star as the central helium abundance 
falls below $Y \sim 0.1$.

To understand which results we might expect in terms of the 
relative duration of the blue phase, hence of the $B/R$ 
ratio, we compare in the panels of fig.~\ref{par_c12_2} 
the main physical and chemical properties of the three models. In the
left-bottom panel we report the variation with time of the effective
temperature: we see that the drop of $T_{eff}$, indicating the return of
the track towards the red, is postponed in the NACRE model, so that the
blue phase lasts $\sim 8$Myr, to be compared to the $\sim 6.5$Myr duration
of the corresponding phase in the CF88 model, and to the
$\sim 7.5$Myr of the Kunz model.  The ratio between the times
spent in the blue and the red region of the diagram is consequently
increase to $t_{blue}/t_{red} \sim 90\%$ ( $\sim 85\%$ in the
Kunz model).

\begin{figure}
\includegraphics[width=8cm]{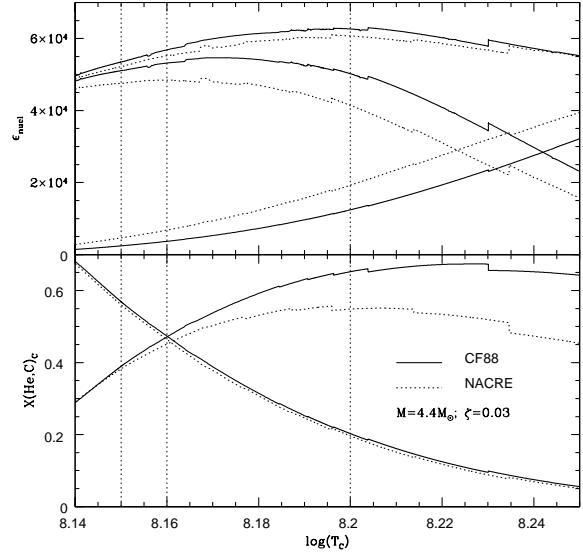}
\caption{{\bf Upper}: The variation with central temperature of the
  coefficients for nuclear energy release during helium burning within
  the interiors of two models presented in fig.~\ref{par_c12_1}.
  The lower lines refer to the energy due to the $^{12}C+\alpha
  \rightarrow ^{16}O$ reaction, the middle lines to the reactions $3\alpha$
  and the upper lines represent the sum of the contributions, giving
  the global energy release. {\bf Lower}: Variation with central 
  temperature of the central abundances of $^4He$ and $^{12}C$
  within the same models.}
   \label{par_c12_4}
\end{figure}

The longer duration of the blue phase is also confirmed by 
the three panels on the right,
which show that the drop of CNO luminosity which is due to the general expansion
of the structure, the phase of the rapid increase of the central density 
following helium exhaustion, and the increase of the central temperature are
all slightly delayed in the NACRE model. The upper and middle panels
on the left show the evolution of the 
chemical abundances of helium and carbon in the
centre of the star. We note that the rate with which helium is consumed is
higher in the CF88 model, while the carbon residual, as expected is larger.
The Kunz models has an intermediate behavior.

To understand the reason of these differences, we focused our attention 
on the physical conditions of the central regions, particularly for what
concerns the thermodynamical properties and the contribution of the two
main reactions to the global nuclear energy release.

First, we report in the two panels of fig.~\ref{par_c12_3} 
the variation of central
temperature and density in the very early phases of the He-burning phase
(note the scale on the abscissa) as a function of the central abundance
of helium. {For clarity reasons, we limited our analysis to the NACRE and CF88
models, the Kunz model showing a behavior very similar to the NACRE.
In both models we note a rapid increase of temperature followed
by a temporary drop, which corresponds to a re-adjustment of the central
regions following the rapid formation of a convective core of approximately 
$M_{core} \sim 0.2M_{\odot}$. The larger values of the NACRE 
$\sigma_{^{12}C+\alpha}$ make the same demand for energy to be satisfied more
easily, so that the temperature for a given helium abundance is slightly
larger in the CF88 model: the differences in temperature in these early 
phases are very small, because the contribution of the quoted reaction to the
global energy release is very poor, but they will increase as
helium burning proceeds. The lower panel shows that no meaningfull differences 
are found in terms of density. On the basis of this, we expect a larger 
energy flux from $3\alpha$ reactions in the CF88 model for a fixed
helium abundance.

In the upper panel of fig.~\ref{par_c12_4} we show 
the variation with central temperature $T_C$ of the coefficient 
for nuclear energy generation $\epsilon_{nucl}$,
together with the separate contributions from the $3\alpha$ and the 
$^{12}C+\alpha$ reactions; the lower panels shows the evolution of the
chemical abundances of helium and carbon. To identify correctly the various
phases, we marked with vertical lines the 
beginning of the excursion to the blue (which starts
in both models when $log(T_C)=8.15$), the bluest point
($log(T_C)=8.16$), and the return to the red, pointing the end of the blue 
phase ($log(T_C)=8.2$). We note that the values of $T_C$ corresponding
to the main evolutive stages are approximately the same for both models; 
therefore, it is the temporal variation of $T_C$ which determines the relative 
duration of the various phases, and the differences between the two
models which we examine here.

We see (fig.~\ref{par_c12_4}) that in both models 
$\epsilon_{3\alpha}$ reaches a maximum when $log(T_C)\sim 8.16$, 
after which the helium depletion dominates
on the temperature increase (we recall that $\epsilon_{3\alpha}
\sim Y^3$). We note a larger $3\alpha$ rate in the CF88 model
which can be understood according to the results shown in the upper 
panel of fig.~\ref{par_c12_3}: 
the CF88 model needs larger temperatures for a given
helium abundance or, equivalently, as we may also see in the bottom panel
of fig.~\ref{par_c12_4}, 
the same temperature corresponds to larger helium abundances,
thus to larger $3\alpha$ rates. The abundance of carbon, as expected, 
is seen in the lower panel to increase faster in the CF88 model, since less
carbon is burnt via the $^{12}C+\alpha \rightarrow ^{16}O$ 
reaction: yet, the energy release
due to this latter reaction is larger (see the upper panel) in the NACRE
model, because of the larger cross-section provided by the NACRE release.
In both models the energy contribution due to $^{12}C+\alpha$ burning is seen
to increase for the whole duration of the He-burning phase, the 
temperature increase more than counterbalancing 
helium depletion (the dependence
of $\epsilon_{^{12}C+\alpha}$ on $X(^4He)$ is weaker with respect to $3\alpha$
reactions, being $\epsilon_{^{12}C+\alpha} \propto X(^4He)X(^{12}C)$); when
$log(T_C)\sim 8.22$, the energy release due to $^{12}C+\alpha$ reaction exceeds
the contribution provided by $3\alpha$ reactions.

The comparison between the global energy release (see the upper panel of 
fig.~\ref{par_c12_4}) shows that for the same central temperature the CF88 model 
attains more nuclear energy by the two major reactions involving
$\alpha$ particles; this determines a more rapid increase of the central
temperatures, speeding-up the whole evolution. This means that the 
same central temperature is attained in the NACRE model somewhat later,
so that the permanence in the blue region, which is terminated for both models
when $log(T_C)\sim 8.2$, is slightly longer.

%

\section{Conclusions}

In this paper we present a detailed study of the helium burning phase
within intermediate mass stars, and we apply our results to the 
LMC cluster NGC1866: we selected this cluster because it shows up 
a well populated clump region, allowing a statistically significant
determination of the colour distribution of stars in the clump. The
observational scenario indicates that the blue region of the clump
is as much populated as the red, with an observed $B/R$ ratio
which is approximately 1.

We tested the influence on the results obtained in terms of expected
$B/R$ ratio of some macro and micro-physics inputs adopted to 
calculate the models, namely: the scheme adopted for mixing 
within nuclear burning regions, the e-folding decay distance of
convective velocities out of the formal border fixed by the
Schwarzschild criterion, the description of convection, and the
rate of the reaction $^{12}C+\alpha \rightarrow ^{16}O$.

Our main findings are the followings:

\begin{enumerate}

\item{The way nuclear burning is coupled to convective mixing within
the central regions during the major phases of nuclear burning
has a strong influence on the results obtained, particularly
in terms of the relative duration of the various evolutive phases.
The models calculated with an instantaneous mixing approximation
burn helium very rapidly, so that the permanence in the blue
region following the overall contraction is shortened by the
subsequent contraction of the nucleus, due to the lack of fuel;
on the contrary, in the diffusive models helium burning is
slower, so that a longer staying in the blue region of the clump
is allowed, in better agreement with the observational scenario.

The differences in terms of the relative duration 
of the blue phase amount approximately to a factor of $\sim 2$.
Thus, our study of NGC1866 suggests that, during core He burning,
mixing in a region of overshooting beyond the convective core boundary
established by the Schwarzschild criterion should not be treated as 
instantaneous, but with some kind of decaying efficiency.

}

\item{The influence of $\zeta$, the parameter determining the e-folding
distance of velocity decay within the radiatively stable regions,
has been carefully tested. Larger values of $\zeta$ allow longer
blue phases; models calculated with a diffusive scheme for mixing
and $\zeta=0.03$ lead to an expected $B/R$ ratio of the order
of $\sim 75\%$.}

\item{The cross-section of the reaction $^{12}C+\alpha \rightarrow ^{16}O$
is the second major source of energy release during helium burning,
so it may change the results obtained in terms of the duration of
various evolutive phases during helium burning. We find that turning
to the NACRE (Angulo et al. 1999) value (which is approximately double
than the value given by the Caughlan \& Fowler (1988) prescription)
leads to increase the permanence in the blue region, so that the
expected percentage of blue stars with respect to the red ones 
exceeds $90\%$, which is in excellent agreement with the observational
scenario. An analogous result is obtained by adopting the cross-sections
for the quoted reaction given by Kunz et al. (2002) .}

\item{The results obtained with the FST and MLT models for turbulent
convection differ only in terms of the colour and the slope
of the giant branch, but not for that concerning the expected $B/R$
ratio. No clear indication regarding convection can be got from
this study.}

\end{enumerate}


\end{document}